\documentclass[showkeys,
reprint,
superscriptaddress,
 amsmath,amssymb,
 aps,
]{revtex4-2}

\usepackage{graphicx}
\usepackage{amsfonts}
\usepackage{xcolor}
\usepackage[breaklinks]{hyperref}
\usepackage{dcolumn}
\usepackage{tabularx}
\usepackage{array}

\begin{document}

\preprint{APS/123-QED}

\title{Superconducting circuits without inductors based on bistable Josephson junctions}


\author{I.~I.~Soloviev}
\email{isol@phys.msu.ru}
\affiliation{Lomonosov Moscow State
	University Skobeltsyn Institute of Nuclear Physics, 119991 Moscow,
	Russia} 
\affiliation{Dukhov All-Russia Research Institute of Automatics, Moscow 101000, Russia}
\affiliation{Physics Department of Moscow State University, 119991 Moscow, Russia}

\author{V.~I.~Ruzhickiy}
\affiliation{Lomonosov Moscow State
	University Skobeltsyn Institute of Nuclear Physics, 119991 Moscow,
	Russia}
\affiliation{Dukhov All-Russia Research Institute of Automatics, Moscow 101000, Russia}
\affiliation{Physics Department of Moscow State
	University, 119991 Moscow, Russia}

\author{S.~V.~Bakurskiy}
\affiliation{Lomonosov Moscow State
	University Skobeltsyn Institute of Nuclear Physics, 119991 Moscow,
	Russia} 
\affiliation{Dukhov All-Russia Research Institute of Automatics, Moscow 101000, Russia}
\affiliation{Moscow Institute of Physics and Technology, State
	University, 141700 Dolgoprudniy, Moscow region, Russia}

\author{N.~V.~Klenov}
\affiliation{Lomonosov Moscow State
	University Skobeltsyn Institute of Nuclear Physics, 119991 Moscow,
	Russia} 
\affiliation{Dukhov All-Russia Research Institute of Automatics, Moscow 101000, Russia}
\affiliation{Physics Department of Moscow State
	University, 119991 Moscow,
	Russia}

\author{M.~Yu.~Kupriyanov}
\affiliation{Lomonosov Moscow State
	University Skobeltsyn Institute of Nuclear Physics, 119991 Moscow,
	Russia} 

\author{A.~A.~Golubov}
\affiliation{Moscow Institute of Physics and Technology, State
	University, 141700 Dolgoprudniy, Moscow region, Russia}
\affiliation{Faculty of Science and Technology and MESA+ Institute of Nanotechnology, 7500 AE Enschede, The Netherlands}

\author{O.~V.~Skryabina}
\affiliation{Lomonosov Moscow State
	University Skobeltsyn Institute of Nuclear Physics, 119991 Moscow,
	Russia}
\affiliation{Moscow Institute of Physics and Technology, State
	University, 141700 Dolgoprudniy, Moscow region, Russia}

\author{V.~S.~Stolyarov}
\affiliation{Moscow Institute of Physics and Technology, State
	University, 141700 Dolgoprudniy, Moscow region, Russia}
\affiliation{Dukhov All-Russia Research Institute of Automatics, Moscow 101000, Russia}

\date{\today}

\begin{abstract}
Magnetic flux quantization in superconductors allows the implementation of fast and energy-efficient digital superconducting circuits. However, the information representation in magnetic flux severely limits their functional density presenting a long-standing problem. Here we introduce a concept of superconducting digital circuits that do not utilize magnetic flux and have no inductors. We argue that neither the use of geometrical nor kinetic inductance is promising for the deep scaling of superconducting circuits. The key idea of our approach is the utilization of bistable Josephson junctions allowing the representation of information in their Josephson energy. Since the proposed circuits are composed of Josephson junctions only, they can be called all-Josephson junction (all-JJ) circuits. We present a methodology for the design of the circuits consisting of conventional and bistable junctions. We analyze the principles of the circuit functioning, ranging from simple logic cells and ending with an 8-bit parallel adder. The utilization of bistable junctions in the all-JJ circuits is promising in the aspects of simplification of schematics and the decrease of the JJ count leading to space-efficiency.
\end{abstract}
\maketitle

\section{Introduction}

The promised end of Moore's law \cite{Moore} nowadays gives rise to the ``beyond Moore's'' technologies. An appropriate alternative to Complementary-Metal-Oxide Semiconductor (CMOS) technology should be fast and scalable, while providing the highest energy efficiency. Superconducting electronics is attractive in this context.

Superconductor technology is known for high clock frequencies, $f_c
\sim 2 - 50$~GHz \cite{EEAD}, and low energy dissipation per logical operation, down to several zJ \cite{zJop}. The advantage over CMOS in energy efficiency reaches two to six orders of magnitude depending on the utilized logic and algorithm \cite{Holmes,Beil,MukhRSFQ,RQL,ASLEDP}. This is especially valuable for an operation at low temperatures. At the standard $T = 4.2$~K, refrigeration cost is $1000\times$ dissipated energy \cite{Holmes,DeBen}. These unique features makes the superconducting circuits to be the most promising candidate for the developing of control electronics of the scalable
computing systems operating across the gradient between room temperature and the temperature of cryogenic payload. They are suitable for the frontier technologies like quantum computers, cognitive radio, scalable sensors, and the quantum internet \cite{IRDS2020}.

At the same time, the integration density of superconducting circuits is far less than current CMOS. The recently demonstrated benchmark circuits for the latest 250~nm and 150~nm MIT LL processes were the shift registers with $7.4 \times 10^6$ and $1.3 \times 10^7$ Josephson junctions (JJs) per square centimeter circuit density, correspondingly \cite{TolpASC2020}. With 4 JJs per bit cell, the functional density is less than 10 Mbit/cm$^2$, showing the need for improvement. Since there is no direct analog of a CMOS transistor in the superconducting element base \cite{Beil}, scaling to higher density requires the development of special approaches. 

\begin{figure}[b]
	\includegraphics[width=1\columnwidth,keepaspectratio]{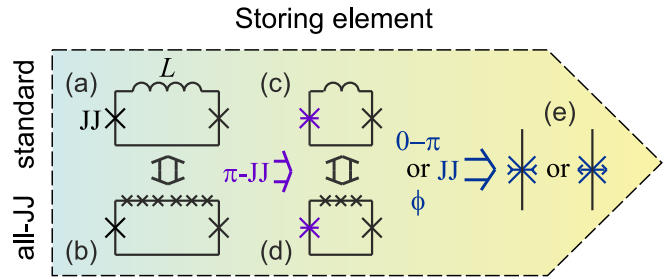}
	\caption{Storing element based on conventional 0-JJs with (a) and without (b) inductor, the one based on 0- and $\pi$-JJ with (c) and without (d) inductor, and bistable $0 - \pi$ and $\phi$-JJ (e). See Fig.~\ref{Fig01} for JJ symbol map.} \label{Fig02}
\end{figure}

The functioning of superconducting logic circuits is based on the effect of magnetic flux quantization, instead of the modulation of conductance \cite{MukhRSFQ,RQL}. This allows a discrete representation of information in the form of a Single Flux Quantum (SFQ), $\Phi_0 = h/2e$ (where $h$ is the Planck constant, $e$ is the electron charge). The basic element of the broadly used Rapid Single Flux Quantum (RSFQ) \cite{RSFQ} logic is a superconducting loop interrupted by Josephson junctions, see Fig.~\ref{Fig02}(a). If the loop inductance, $L$, is high enough such that $I_c L \approx
\Phi_0$ (where $I_c$ is the JJ critical current) then an SFQ can be hold in the loop representing logical unity, while an absence of an SFQ means logical zero. 

In this paper, we consider different approaches for miniaturization of RSFQ basic cell. This includes the analysis of the scalability of the inductor, the substitution of the inductor by Josephson junctions, and the utilization of magnetic JJs of various kinds. The latter ends with the possibility of replacing the basic cell with a single bistable Josephson junction (see Fig.~\ref{Fig02}). We complement this with a proposition of a design methodology for superconducting circuits without inductors based on bistable JJs. Our design is based on a basic block. Modification and complication of its schematic allow obtaining various logic cells. We demonstrate the validity of the proposed methodology using the design of an 8-bit parallel adder as an example. We conclude the article with a brief discussion of the obtained results and outlining further research directions.

\section{Miniaturization of basic cell}

\subsection{Scaling of inductor}

Theoretical estimation \cite{TolpKinInd} of the maximum density of SFQ-based circuits utilizing geometrical inductance of wires gives already achieved $\sim 10^7$~JJ/cm$^2$. The further decrease of the linewidth and spacing is problematic because of a nearly exponential growth in the mutual inductance and cross-talk between the inductors \cite{KinInd}. The main approach to shrink the inductor is related to the utilization of kinetic inductance. Here the energy stored in the inductor corresponds to the kinetic energy of the superconducting current, not to the magnetic field around a wire. The critical current of the inductor, $I_{ck}$, must be greater than that of a Josephson junction, which in turn should be much greater than the noise current, $I_{ck} > I_c \gg I_n = (2\pi/\Phi_0)k_BT$ (where $k_B$ is the Boltzmann constant), to provide low bit error rate. The noise current at the operation temperature $T = 4.2$~K is $I_n \approx 0.18~\mu$A. The typical value of the Josephson junction critical current is $I_c \sim 0.1$~mA so that the inductance of a storing cell is  $L_{cell} \approx \Phi_0/I_c \sim 20$~pH.

For a real type-II superconductor film with the width, $w$, being much larger than the coherence length, $w \gg \xi$, the wire critical current, $I_{ck}$, is a fraction of Ginzburg-Landau (GL) depairing critical current, $I_{ck} \approx \eta I_{GL} = wd\eta\Phi_0/(3\sqrt{3}\xi\mu_0\lambda^2)$ \cite{GL,KLPB}. Here, $d$ is the thickness of the wire, $\lambda$ is the London magnetic penetration depth, and $\mu_0$ is the vacuum permeability. The critical current reduction \cite{KinInd}, $\eta \sim 0.1$, is caused by the entry and motion of Abrikosov vortices from the film edges.

The kinetic inductance of a thin wire is $L_k = l\mu_0\lambda^2/(wd)$, where $l$ is the wire length, $d \ll \lambda$ . The required cell inductance, $L_{cell} = L_k \sim l/(wd)$. The critical current of the inductor must also be greater than the JJ, so $I_c < I_{ck} \sim wd$. Therefore, the minimum wire length and its cross-section area are determined by the chosen material and the quality of the wire edges. For a reasonable ratio, $I_{ck}/I_c \geq 4$, and some ``dirty'' superconductor like MoNx, NbTiN or NbN with high kinetic inductance (where $\lambda \approx 500$~nm, $\xi \approx 10$~nm, $\xi/\eta = 44$~nm \cite{KinInd}), the wire geometrical parameters are $l = (I_{ck}/I_c)3\sqrt{3}\xi/\eta \geq 0.9~\mu$m and $wd = l\mu_0\lambda^2/L_{cell} \geq 14000$~nm$^2$.

If the wire width is equal to the minimum feature size of a standard SFQ5ee MIT LL process technology, $w = 350$~nm, the wire thickness is $d \geq 40$~nm so that the wire cross-section aspect ratio is $w/d \leq 8.75$. The area of the kinetic inductor is $wl \geq 0.32~\mu$m$^2$, which is close to the typical area of a Josephson junction, $\pi w^2 \approx 0.4~\mu$m$^2$.  The inductor implementation requires about 2.5 squares with the inductance per square, $L_{k\Box} \leq 8$~pH. The obtained numbers show that the utilization of the kinetic inductance is a good solution for the current technological processes. 

However, the convenience of utilization of the kinetic inductance at smaller scales is questionable. For example, the reduction of the inductor width down to $w = 90$~nm requires the increase in its thickness up to $d \approx 160$~nm (if the length is $l = 0.9$~$\mu$m) to preserve its critical current. Here the wire cross-section aspect ratio is $w/d \approx 0.56$ that presents a difficulty for fabrication. At the same time, the inductance per square becomes four times smaller so that the implementation of the inductor is much less effective in the new scale. A safe threshold for the kinetic inductor scaling is about $w = (I_{ck} 3 \sqrt{3} \xi \mu_0 \lambda^2/\eta\Phi_0)^{1/2} \approx 120$~nm, which provides a square cross-section of the wire. Based on the presented estimations, we conclude that the basic cell scaling requires further research of alternative methods.

\subsection{Superconducting loop without inductor}

\begin{figure}[t]
	\includegraphics[width=1\columnwidth,keepaspectratio]{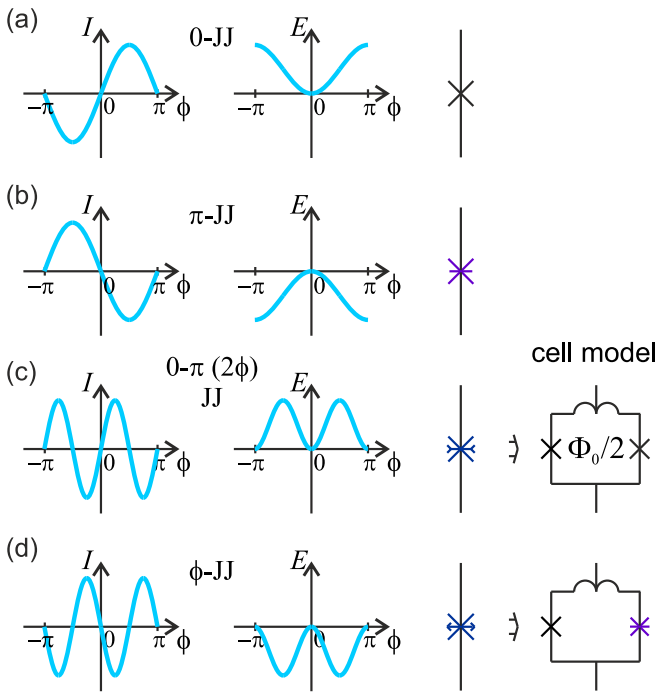}
	\caption{Current-phase relation, $I(\phi)$, energy-phase relation, $E(\phi)$, and symbol of 0-JJ (a), $\pi$-JJ (b), $0-\pi$ JJ (c), and $\phi$-JJ (d). CPR and EPR of $0-\pi$ and $\phi$-JJ are presented for the case of the suppressed first harmonic of their CPR. Cell models are presented for $0-\pi$ ($2\phi$) and $\phi$-JJ.
   } \label{Fig01}
\end{figure}

If an SFQ is inside the standard cell shown in  Fig.~\ref{Fig02}(a), the Josephson phase of one of the junctions is close to $2\pi n$ (where $n$ is an integer) while the phase of another one is about $2\pi (n-1)$, so that the total phase increment in the circuit is $2\pi$. These Josephson phase values correspond to minima of conventional superconductor - insulator - superconductor JJ (0-JJ hereinafter) Energy-Phase Relation (EPR), $E/E_{J0} = 1 - cos(\phi)$, where $\phi$ is the Josephson phase, and $E_{J0} = I_c \Phi_0/2 \pi$ is the Josephson energy, see Fig.~\ref{Fig01}(a). This EPR corresponds to a sinusoidal Current-Phase Relation (CPR), $I = I_c \sin(\phi)$, where $I$ is the current flowing through the junction. 

Josephson junctions in the storing cell can be connected by a stack of JJs instead of an inductor \cite{Tolp,StJJ}, see Fig.~\ref{Fig02}(b). Such cell can be called all Josephson junction (all-JJ) circuit. If the critical currents of the JJs in a stack are identical, the phase drop on each of them is $2\pi/N$, where $N$ is the number of the JJs in the stack. To prevent the JJs from switching, this phase drop should be less than $\pi/2$ (in accordance with 0-JJ CPR). Reliable operation of circuits is supposed to be provided with the phase drop about $\pi/3$ \cite{Tolp}. This gives 6 JJs in a stack that can be implemented, e.g., using two 3-JJ stacks \cite{StJJ}.

\subsection{$\pi$ Josephson junction}

The issue of a rather large number of JJs in a stack can be mitigated by the introduction of the JJ having ferromagnetic interlayer in its weak-link region providing $\pi$ phase shift of its CPR, $I = I_c \sin(\phi + \pi)$ ($\pi$-JJ hereinafter) \cite{RyazPiJJ,GKI}. $\pi$-JJ CPR and EPR are presented in Fig.~\ref{Fig01}(b). The $\pi$-JJ symbol is a cross with a dash in the middle. With a $\pi$-JJ in a loop, the phase increment corresponding to magnetic half flux quantum already exists in the cell. The cell inductance can be reduced correspondingly, and the number of JJs in the substituting stack can be decreased down to 3, see Fig.~\ref{Fig02}(c),(d) \cite{TFF1,TFF2,TFF3,PJORT,piJJDR,Yo2,HFQ1,HFQCir,HFQ2}. In RSFQ Toggle (T) flip-flop, it is possible even to completely substitute the inductor by a $\pi$-JJ \cite{TFF1,TFF2,TFF3}. Unfortunately, this can't be a general solution for all circuits. The constraints of this method come from the persistent current arising due to the $\pi$-JJ inherent phase shift. This current outflows into the neighboring cells and may alter their power supply and break their operation. The direction of this current is determined by the circuit state and thus can't be compensated with constant bias current adjustment. 

\begin{figure}[t]
	\includegraphics[width=1\columnwidth,keepaspectratio]{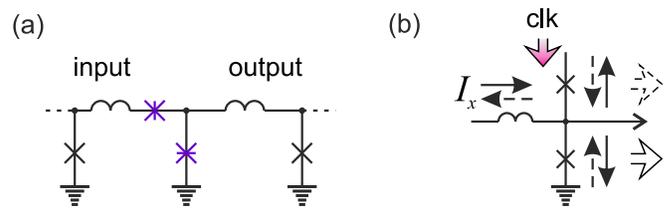}
	\caption{(a) Two-loop circuit with $\pi$-JJs. (b) Balanced comparator \cite{RSFQ}.} \label{Fig0}
\end{figure}

A design methodology utilizing two $\pi$-JJs connected in series was proposed to circumvent this issue \cite{piJJDR}. Fig.~\ref{Fig0}(a) illustrates the idea. Here the Josephson phase of each of the $\pi$-JJs is about $\pi + 2\pi n$ while the one on the 0-JJ in the input circuit is about $2\pi n$; combined, the phase shift is $2\pi n$, allowing a state with $n=0$ and thus zero phase shift and zero current (see also Fig.~\ref{Fig01}(a),(b)). This allows the safe connection of the circuit input to conventional RSFQ cells. At the same time, $\pi + 2\pi n$ phase drop on the central $\pi$-JJ makes 0-JJ of the output loop being outside the minimum of its potential energy. Minimization of the total energy of the $\pi$- and 0-JJ leads to the states with a noticeable current in the output loop, which can have opposite directions. This provides the opportunity to read out the circuit state by using conventinal balanced comparator shown in Fig.~\ref{Fig0}(b) \cite{RSFQ}. The latter is two JJs connected in series, one of which is switched by the clock (clk) pulse depending on the direction of the measured current, $I_x$. Many broadly used cells like a Delay (D) flip-flop and Non-Destructive Read-Out (NDRO) \cite{piJJDR}, logical AND, and OR cells \cite{Yo2} are designed in the frame of this approach. 

\subsection{Bistable Josephson junctions}

The main requirement for a storing cell is the existence of two stable states: with and without an SFQ inside. However, the desired bistability can be obtained even with a single bistable Josephson junction. In this case, the storing element of the circuits reduces to this single JJ, and its state doesn't relate to an SFQ. 


The bistability corresponds to the existence of the second harmonic in the JJ CPR,  $I = A\sin(\phi) + B\sin(2\phi)$. 
If $|B| > A/2$ , the EPR has a double-well shape \cite{Suppl}. In the case of the positive second harmonic amplitude, $B > 0$, the JJ energy minima are located at 0 and $\pi$ phase values \cite{Suppl}. This junction is called $0-\pi$ JJ, correspondingly \cite{0piJJ1,0piJJ2,0piAPhiJJ}. Note that the vanishing of the first harmonic here leads to the doubling of the CPR. In this case, the JJ is called $2\phi$-JJ  \cite{T,RHM,PBRB,BSPR,Ryaz2phi}. If the second harmonic amplitude is negative, $B < 0$, the JJ energy minima is symmetrically located around zero at $\pm \phi$ phases, where $|\phi| \leq \pi/2$ \cite{Suppl}. This JJ is called $\phi$-JJ \cite{PhiJJ1,PhiJJ2,PhiJJ3,PhiJJ4mem,PhiJJ5,0piAPhiJJ}. CPR and EPR of $0-\pi$ and $\phi$-JJ for the case of suppressed first harmonic in their CPR are presented in Fig.~\ref{Fig01}(c),(d). The symbols of the JJs are similar to the one of the $\pi$-JJ but with arrows added in the ends of the central dash directed inside or outside for $0-\pi$ and $\phi$-JJ, correspondingly.

The considered bistable junctions are relatively new types of JJs. They are not introduced in digital superconducting technology yet. While the optimal method for their fabrication is a technological challenge, they can be already modeled using cells composed of standard junctions \cite{PhiJJSQ}. The model of $2\phi$-JJ is a superconducting loop with two 0-JJs and a Half Flux Quantum (HFQ) applied to the cell, see  Fig.~\ref{Fig01}(c) \cite{Suppl}. The cell has a small but finite inductance. The smaller the inductance, the smaller is the cell critical current and the closer is the cell CPR to the doubled sine. Here the decrease of the first harmonic amplitude is caused by the circulating current induced by HFQ, while the second harmonic amplitude, and hence the effective critical current of the junction, is determined by the inductance.

In a very similar way, a $\phi$-JJ can be modeled by a cell with 0- and $\pi$-JJ, see  Fig.~\ref{Fig01}(d) \cite{Suppl}. It was shown that Josephson Transmission Line (JTL) composed of such cells (modeling $\phi$-JJs) is capable to transmit HFQ \cite{HFQ1,HFQCir,HFQ2}. The transfer to manipulation with HFQ instead of SFQ makes the HFQ circuits to be more power-efficient than all other superconducting logic circuits \cite{HFQ2}. Note that the small inductance in the cell model can be substituted by 0-JJs to obtain all-JJ circuit \cite{Suppl}.

The utilization of bistable JJs seems promising in the aspects of power and space efficiency. Below we propose a design methodology for the circuits without inductors based on bistable $2\phi$-JJs. The latter are chosen because their cell model contains only conventional 0-JJs, see Fig.~\ref{Fig01}(c). Since the proposed design requires only two types of JJs, the conventional 0-JJs and bistable $2\phi$-JJs, this simplifies the experimental verification of prototype circuits. The information bit in the circuits is represented as a presence or absence of a $2\pi$ superconducting phase change. This phase change can be transferred along the circuit by the application of the bias current.

\section{Design methodology}

\subsection{Basic block}

The dynamics of the $2\phi$-JJ, as well as 0-JJ, is described by the well-known Resistive Shunted Junction model with Capacitance (RSJC) \cite{RSJC} (see also \cite{Suppl} for details):
\begin{equation} \label{RSJ}
i = A \sin\phi + B \sin 2\phi + \alpha\dot{\phi} + \ddot{\phi},
\end{equation}
where current flowing through the junction, $i$, and the amplitudes of CPR harmonics, $A, B$, are normalized to the critical current of a reference junction, $I_c$, dots mean derivatives with respect to the time, $t$, normalized to the inverse plasma frequency, $\tau = t\omega_p$, which is determined by the constants of a certain fabrication process, $\omega_p = \sqrt{2\pi j_c/\Phi_0 c}$, where $j_c$ is the critical current density of Josephson junctions and $c$ is their specific capacitance, $\alpha = \omega_p/\omega_c$ is the Josephson junction damping coefficient, and $\omega_c = 2\pi I_c R/\Phi_0$ is the junction characteristic frequency, $R$ is the junction resistance in the normal state. 

Each type of the considered junctions has only one harmonic dominating in the CPR. Therefore, the dominating harmonic determines the critical current of the junction. Since the circuits consist of the JJs only and each JJ has only two parameters, the critical current and the damping coefficient \cite{Suppl}, these parameters determine the circuit dynamics. The former reflects the strength of superconducting coupling determining the potential barrier for the JJ switching, while the latter reflects the duration of the switching.

\begin{figure}[t]
\includegraphics[width=1\columnwidth,keepaspectratio]{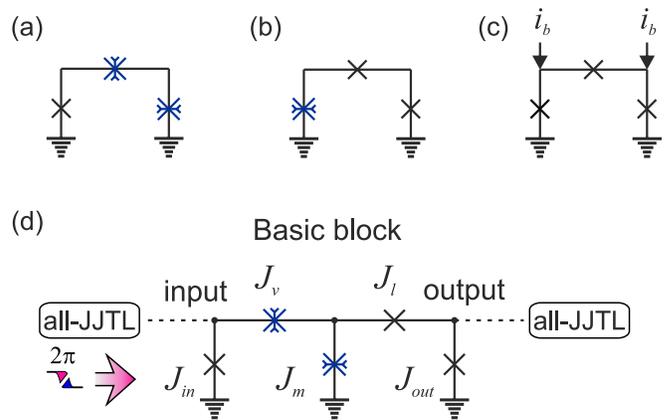}
\caption{Interface circuits to change (a) and read out (b) $2\phi$-JJ state. (c) all-JJTL basic cell. (d) Schematic of the basic block with interface all-JJTLs.} \label{Fig1}
\end{figure}

The proposed circuits are based on the basic block designed in accordance with the methodology presented in \cite{piJJDR}: the input loop is in a current-less state, while the different phases of the $2\phi$-JJ correspond to the different currents circulating in the output loop.

The input loop is composed from one 0- and two $2\phi$-JJs, see Fig.~\ref{Fig1}(a). The $2\pi n$ phase drop on the 0-JJ is distributed among $2\phi$-JJs meaning the phase drop multiple of $\pi$ on each of them. This corresponds to the current-less state of the loop due to CPRs of the used JJs, see Fig.~\ref{Fig01}. Therefore the loop can be connected to other circuits from the side of 0-JJ. 

A circulating current in the circuit with two 0-JJs and one $2\phi$-JJ, see Fig.~\ref{Fig1}(b), corresponds to $\pi + 2\pi n$ phase drop on the latter one. At the same time, $2\pi n$ phase drop on $2\phi$-JJ provides the current-less state. Combination of the two considered loops (Fig.~\ref{Fig1}(a),(b)) forms our basic block, see Fig.~\ref{Fig1}(d).

In our simulations, we utilize all-JJTLs to apply and read out superconducting $2\pi$ phase changes to/from the circuits. These are the JTLs in which each inductor is substituted for a single Josephson junction, see all-JJTL cell in Fig.~\ref{Fig1}(c). Parameters of the circuits, as well as a detailed description of their dynamics, are presented in supplemental materials \cite{Suppl}.

\begin{figure}[]
	\includegraphics[width=1\columnwidth,keepaspectratio]{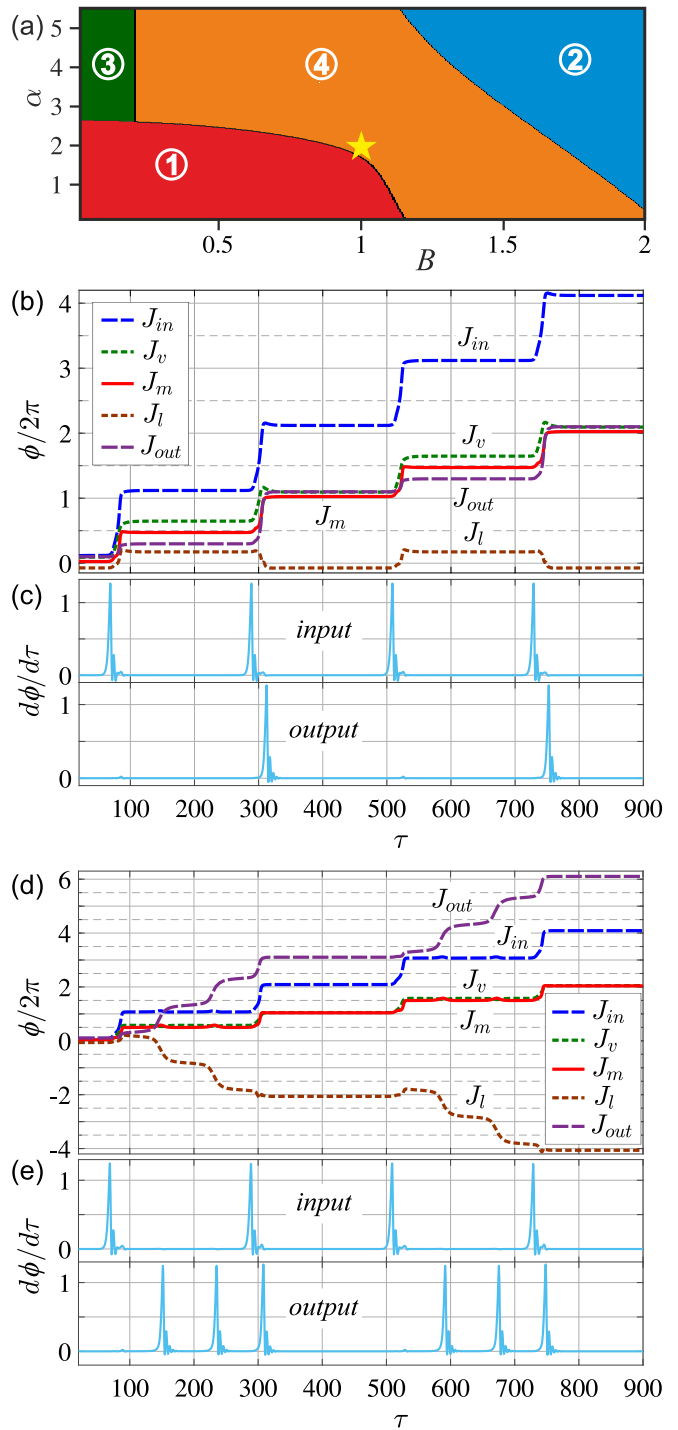}
	\caption{(a) The input valve junction $J_v$ (shown in Fig.\ref{Fig1}(d)) parameters ($B,
		\alpha$) corresponding to different modes of the basic block operation: 1 - terminator, 2 - transmission line, 3 - digital frequency divider, and 4 - oscillator. The main junction $J_m$ parameters are depicted by a star ($\star$). (b) Dynamics of phases of the Josephson junctions and (c) voltage pulses ($d\phi/d\tau$) corresponding to the phase change waves in the input and output all-JJTLs in mode 3 for $B_v = 0.1, \alpha_v = 5$ . (d), (e) The same, correspondingly, for mode 4 ($B_v = 0.5, \alpha_v = 5$).} \label{Fig2}
\end{figure}

The logical state of the basic block corresponds to the Josephson phase of the main $2\phi$-JJ, $J_m$, located in the center of the circuit (Fig.~\ref{Fig1}(d)). $J_m$ parameters are depicted on ($B, \alpha$) plane in Fig.~\ref{Fig2}(a) by a star as a reference point. The superconducting phase change wave enters the basic block through the junction $J_{in}$ and can exit through $J_{out}$. The interconnecting junctions, $J_v$ and $J_l$, serve as the input and output valves in the wave transfer process.

The block possesses four modes of operation in dependence of the input valve junction, $J_v$, parameters ($B, \alpha$) as it is presented in Fig.~\ref{Fig2}(a): (1) - terminator, (2) transmission line, (3) digital frequency divider, and (4) oscillator. In a trivial case where the junction $J_v$ is sufficiently ``weaker'' and ``faster'' than the main junction $J_m$, the input $2\pi$ phase change switches $J_{in}$ and exits through $J_v$ (mode 1). In the opposite case, the input wave passes through $J_m$ switching it twice and exits through $J_{out}$ (mode 2).

\begin{figure*}[]
	\includegraphics[width=1\textwidth,keepaspectratio]{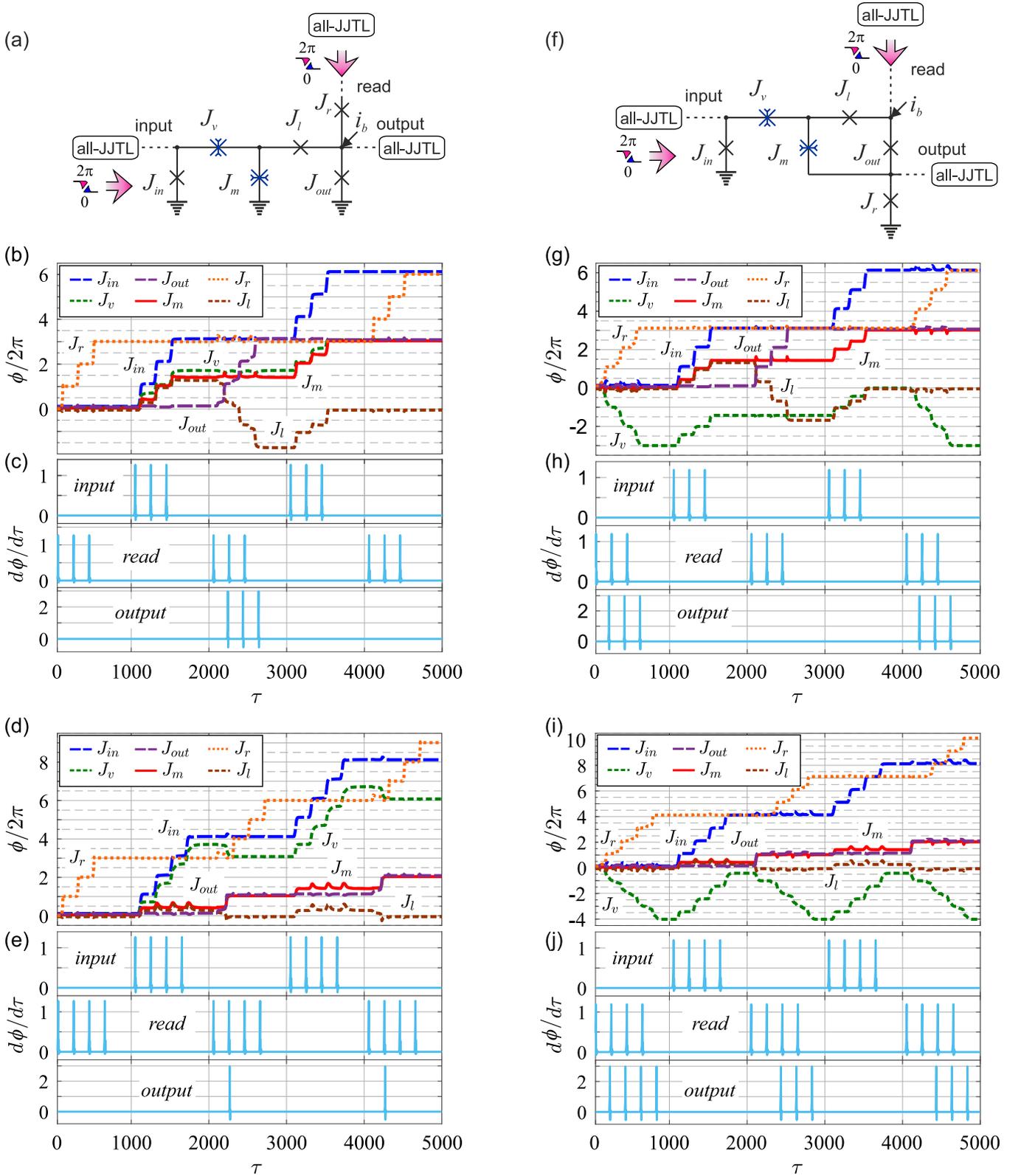}
	\caption{(a) All-JJ cells based on the basic block with controlled readout (NDRO and DRO) of the main $2\phi$-JJ states without (a) and with (f) inversion of the read data. Dynamics of phases of the Josephson junctions and voltage pulses corresponding to the propagation of the phase change waves ($d\phi/d\tau$) in the input, read and output all-JJTLs for NDRO ((b), (c)), DRO ((d), (e)), NDRO with output inversion ((g), (h)) and DRO with 		output inversion ((i), (j)) cells. Parameters of the Josephson junctions are presented in Table~I in \cite{Suppl}.} \label{Fig3}
\end{figure*}

More interesting modes of operation are obtained at the mixed ratio of parameters of the $2\phi$-JJs. If the input valve junction is ``weaker'' than the main junction it starts to switch first. But if its switching takes a relatively long time, the main junction also has time to switch (see Fig.~\ref{Fig2}(b)). In this way, the incoming $2\pi$ phase change is transformed into $\pi$ phase increase on each of the $2\phi$-JJs. Since the output junction $J_{out}$ has a $2\pi$ periodic CPR, every second $2\pi$ phase change wave passes through the basic block. Here the block operates as a digital frequency divider like a T flip-flop (mode 3), see Fig.~\ref{Fig2}(b),(c).

Simultaneous switching of the $2\phi$-JJs occurs in a relatively wide range of ratios of their parameters (including the inverse ratio with the  corresponding swap in the sequence of their switchings). Interestingly, increase of the input valve junction critical current results in another operation regime. Here leakage of the interface all-JJTL bias current into $J_{out}$ provides its periodic switching if $J_m$ phase is about $\pi + 2\pi n$. These switchings of $J_{out}$ can be turned on and off by the switching of the logic state of the basic block. In this mode 4, the block operates as a conventional SFQ-to-DC converter \cite{RSFQ}., see Fig.~\ref{Fig2}(d),(e).

\subsection{Controlled readout of $2\phi$-JJ states}

Controlled readout of the circuit logical state can be performed if the output of the basic block is designed as a balanced comparator (see Fig.~\ref{Fig0}(b) \cite{RSFQ}). In this case, another 0-JJ, $J_r$, is connected to the output loop, accordingly, see Fig.~\ref{Fig3}(a). Read $2\pi$ phase change wave is applied to the pair of junctions, $J_r$, $J_{out}$.

The Non-Destructive Read-Out (NDRO) is realized when the processes in the output loop do not significantly affect $J_m$. At the same time, the $J_m$ phase is increased by about $\pi$ with every $2\pi$ phase change wave coming from the input loop. The circuit operation is close to the modes 3, 4 considered above but here the input wave never passes to the output. The direct passages of the waves from the input to the output and backward are blocked by making the output valve junction, $J_l$, ``weak'' and ``fast'' compared to the neighboring junctions, $J_m$, $J_{out}$. Dynamics of Josephson junctions and voltage pulses corresponding to the propagation of the phase change waves in the input, read and output all-JJTLs are presented in Fig.~\ref{Fig3}(b),(c). 

One can switches from NDRO operation to the Destructive Read-Out (DRO) by changing the output valve junction, $J_l$, parameters; namely, making it slower. The DRO is implemented when the only first input $2\pi$ phase change wave increases $J_m$ phase by about $\pi$. Another $\pi$ increase is available only with the read wave. Since $J_m$ is connected to $J_l$, the slower phase increase in the point of their connection allows $J_v$ to switch twice with every input $2\pi$ phase change wave if $J_m$ phase is about $\pi + 2\pi n$, see \cite{Suppl}. At the same time, with a slowdown of $J_l$, $J_m$ has time to switch from the $\pi + 2\pi n$ state while $J_{out}$ phase increases by $2\pi$. The DRO cell operates as a D flip-flop, see Fig.~\ref{Fig3}(d),(e).

Note that the read data can be easily inverted. This is achieved by a simple swap of $J_r$, $J_{out}$ connection order, see Fig.~\ref{Fig3}(f). In this case, $J_m$ and $J_{out}$ are connected to the ground through $J_r$.
Thus, the stationary distribution of the currents and dynamics of the junctions become changed so that the circuit parameters require an additional adjustment, see \cite{Suppl}. Simulation of dynamics of NDRO and DRO cells with output inversion is presented in Fig.~\ref{Fig3}(g),(h) and Fig.~\ref{Fig3}(i),(j), correspondingly. The DRO cell here performs synchronous data inversion that is the logical NOT operation.

\subsection{Example of a cell design}

\begin{figure}[!t]
\includegraphics[width=1\columnwidth,keepaspectratio]{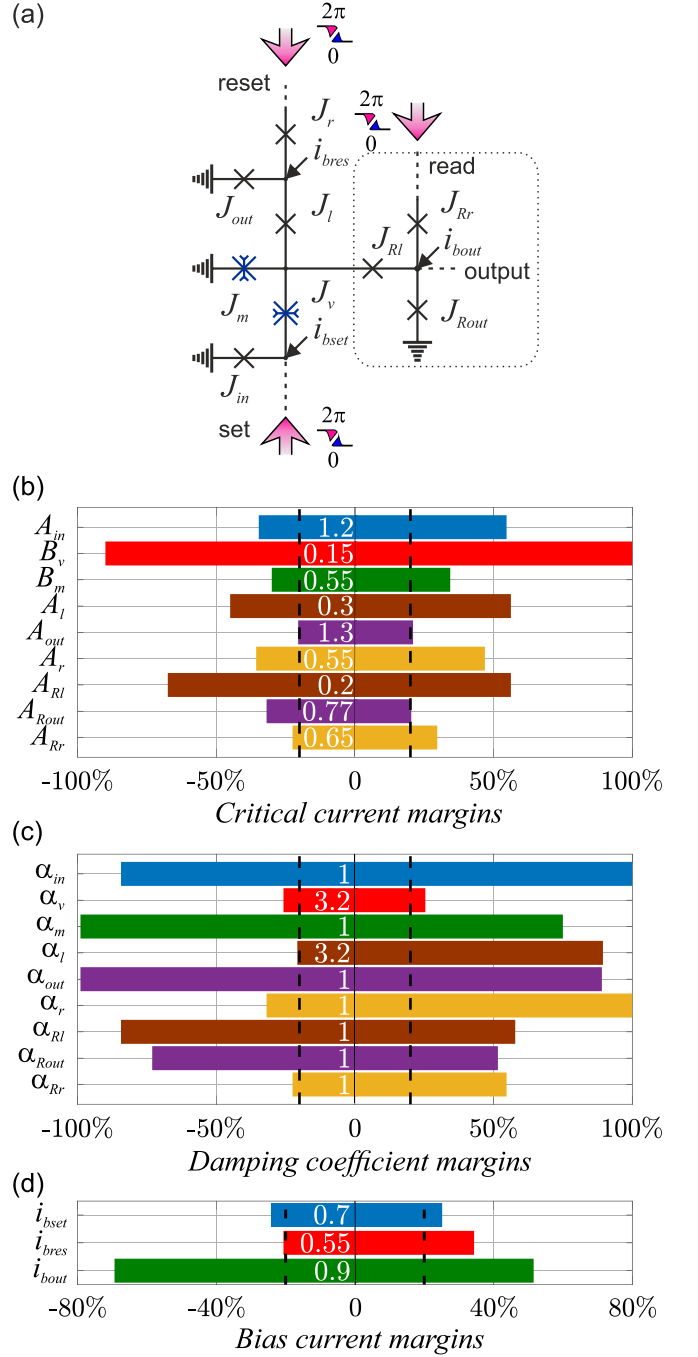}
\caption{(a) Schematic of NDRO cell with separate inputs to set/reset
the latch formed as a combination of DRO cell with additional NDRO branch
(the last one is circled by a dotted line). Working margins of critical
currents (b), damping coefficients (c) and bias currents (d). Optimal values of parameters are shown in corresponding
panels. $\pm 20\%$ boundaries are shown by vertical dotted lines.} \label{Fig4}
\end{figure}

We presented only a simple combination of the input and output circuits to write/read the logical state encoded in a $2\phi$-JJ phase so far. Functionality of a cell can be increased by connection of additional branches. For example, NDRO output jointed with DRO cell forms NDRO cell with separate inputs to set/reset the latch, see Fig.~\ref{Fig4}(a). Additional output interface circuit providing NDRO is circled by a dotted line. Josephson junctions of this circuit are marked by capital ``R'' in their subscripts. Designation of the other Josephson junctions are inherited from Fig.~\ref{Fig3}(a). $2\pi$ phase change wave entering the cell through $J_{in}$ switches the logical state by increasing the $J_m$ phase by $\pi$, while the one entering the cell through $J_r$ resets the state back. The critical current of the Josephson junction connecting NDRO branch to DRO cell is made small to make these parts decoupled.

In the design of this cell, we divide all the JJs into ``fast'' and ``slow'' ones, having a correspondingly small or large value of the damping coefficient, see Fig.~\ref{Fig4}(c). After optimization procedure, we find that the working margins of the bias currents, as well as of parameters of the Josephson junctions, are greater than $\pm 20\%$ at low-speed test, see Fig.~\ref{Fig4}(b)-(d). The margins are expected to shrink with the operation frequency increase.

\begin{figure*}[]
	\includegraphics[width=1\textwidth,keepaspectratio]{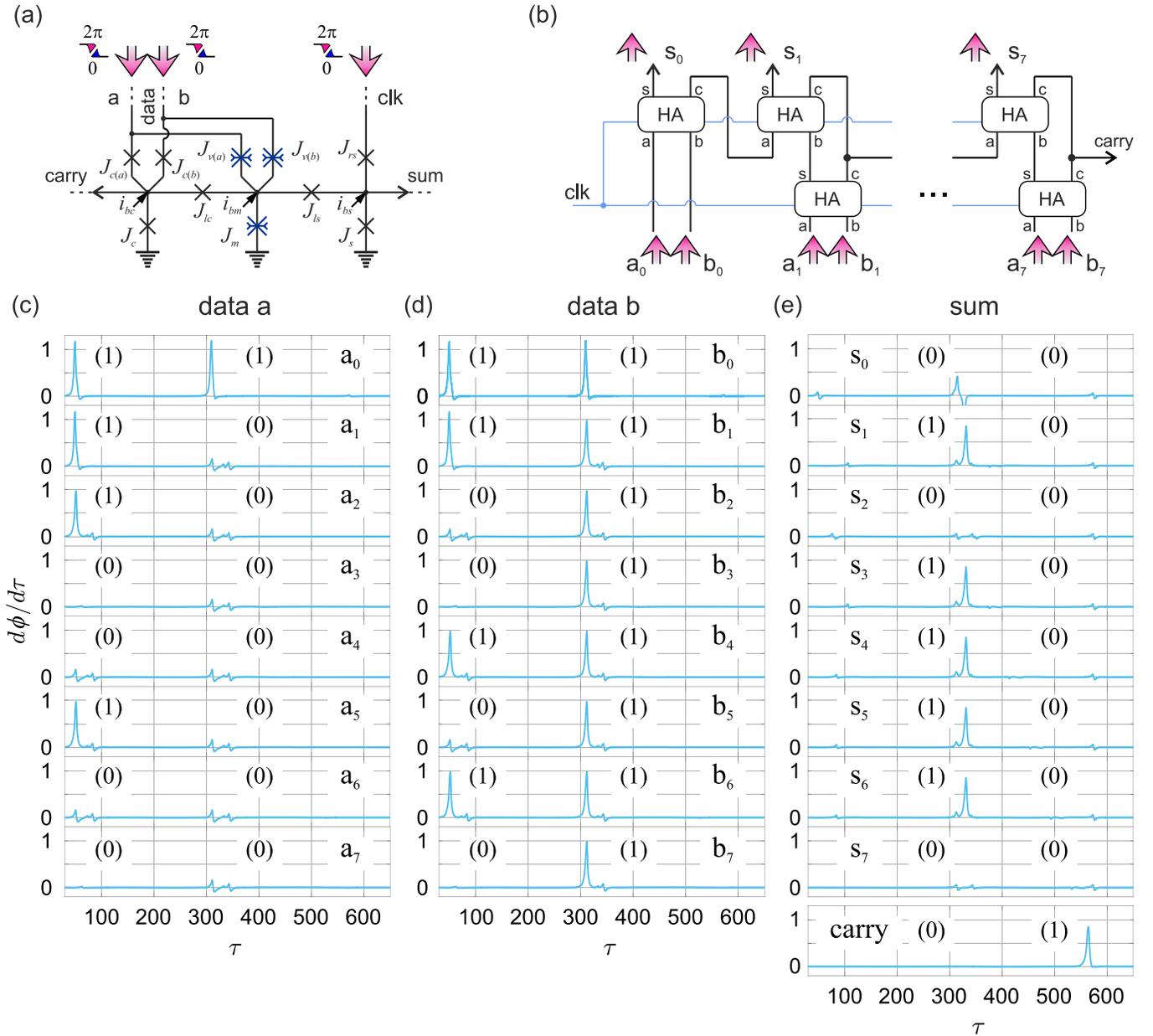}
	\caption{(a) All-JJ half-adder schematic. (b) 8-bit ripple-carry adder schematic. Voltage pulses corresponding to phase change waves in the data branches $a$ (c) and $b$ (d), and the sum branches (e) with the last bottom panel in the column presenting the output carry.} \label{Fig5}
\end{figure*}

We add a small parasitic inductance in series to every Josephson junction of the considered circuit to examine its effect. The normalized inductance value is $l = 2\pi I_c L/\Phi_0 = 0.5$ (1.65~pH for $I_c = 0.1$~mA). While the values of parameters become clearly shifted, after additional optimization procedure the working margins are restored to upper than $\pm 20\%$. This illustrates the importance of careful extraction of parameters of the circuits from layout as well as overall robustness of the proposed design approach to existence of the parasitic inductors.

\subsection{Benchmark logic circuit design}


We verify the validity of the proposed design methodology on the example of the design of an 8-bit parallel adder. The basic element of this adder is a half-adder (HA) presented in Fig.~\ref{Fig5}(a). Here the phase of the main $2\phi$-JJ, $J_m$, is changed by every $2\pi$ phase change wave coming from the data lines $a$ or $b$ through $J_{v(a,b)}$ junctions. The interface circuit of the output sum ($J_{ls}, J_{rs}, J_s$) provides destructive readout of the $J_m$ state. The interface circuit of the output carry ($J_{lc}, J_{c(a,b)}, J_c$) provides a non-destructive readout of the $J_m$ state before it is changed by the data phase change waves. Therefore, the carry readout is asynchronous with respect to clock. Note that data can flow sequentially or simultaneously to the HA cell. In the last case the output junction of the carry branch, $J_c$, is switched due to the total effect of the two $2\pi$ phase change waves. This is similar to the operation of RSFQ AND cell \cite{RSFQ}. Parameters of the half-adder JJs are presented in Table~II in \cite{Suppl}.

Note that just a single $2\phi$-JJ is sufficient here to provide the data storage. This is in contrast to RSFQ HA where insensitivity to the data delay comes at the cost of doubling the SFQ storage circuit \cite{RSFQHA}. Thus, the number of JJs in the presented schematic is even smaller than in the RSFQ counterpart despite the substitution of inductors for JJs.

An 8-bit ripple-carry wave pipeline adder schematic is presented in Fig.~\ref{Fig5}(b). Its energy-efficient RSFQ (ERSFQ \cite{ERSFQ}) counterpart was proposed earlier as a process benchmark \cite{ER8bAdd}. For $N$-bit circuit one needs $2N - 1$ HA cells connected in our case by all-JJTLs and all-JJ confluence buffers (CBs). The total number of JJs in the adder is 362 with 150 JJs in HAs (10 JJs per HA), and 212 JJs in all-JJTLs and CBs. An additional 150 JJs are in the clock distribution network. The summation operation is performed in a single clock cycle \cite{ER8bAdd} with the total time delay (normalized to the inverse plasma frequency), $\tau_\Sigma = 275$. The output data front is aligned, meaning that the less significant and the most significant bits are produced simultaneously. 

Circuit operation was simulated with various inputs. Voltage pulses corresponding to the phase change waves in $a$ and $b$ data branches as well as in the sum branches are shown in Fig.~\ref{Fig5}(c),(d),(e), correspondingly. The last bottom panel in Fig.~\ref{Fig5}(e) shows the output carry. The first example presents the summation of two randomly chosen numbers: $a = 0010~0111$ (39) and $b = 0101~0011$ (83) that results in the sum~$= 0111~1010$ (122). The second example corresponds to the longest propagation of carry: $a = 0000~0001$ (1) and $b = 1111~1111$ (255) that results in the sum~$= 0000~0000$ and generation of the output carry (256).

\section{Discussion}

With the presented utilization of $2\phi$-JJs one can easily mimic the behaviour of RSFQ circuits due to a certain $2\phi$-JJ CPR. However, other types of bistable junctions are also suitable. Using 0-$\pi$ JJs instead of $2\phi$-JJs in the basic block would result in some shrinking of operation margins caused by lesser depth of the potential well at $\phi = \pi$. Using $\phi$-JJs with potential energy minima at $|\phi|  \leq \pi/2$ would additionally lead to some finite current in the output interface circuit in both states of the $\phi$-JJ. Still, correct circuit operation is possible using these alternative implementations \cite{piJJDR}.

A nonzero first harmonic in the CPR provides two critical currents of the bistable JJ. This is due to the different heights of potential barriers between the EPR minima. The difference in the critical currents gives an opportunity to read out the state directly from the junction \cite{PhiJJ4mem}. This could serve for further simplification of the schematics.

Logical zero and unity correspond to zero and $2\pi$ phase changes in the presented circuits. At the same time, the operation of the cells is based on the transformation of the $2\pi$ phase change into $\pi$ phase drops on the bistable JJs. One can consider an option for a more complex information processing with propagation of $\pi$ phase change waves. This includes a possibility of ternary logic circuit design. Three options of the logical states can be associated with zero, $\pi$, and $2\pi$ phase changes. Manipulation with $\pi$ phase change waves instead of $2\pi$ ones can additionally increase the power-efficiency \cite{HFQ2}. 

The utilization of bistable JJs eliminates the need for SFQ storage and hence the necessity for technological implementation of the quantizing inductance. The inductor can be used as an interconnecting element or not used at all. The latter leads to the fact that the circuit characteristics are determined only by the parameters of the Josephson junctions. Our results show that the all-JJ circuits can be designed using just a few values of JJ damping, which is promising for their implementation. While circuits without inductors can be implemented with more conventional 0- and $\pi$-JJs, utilization of bistable Josephson junctions significantly decreases the total JJ count. 

We have shown that the all-JJ cells can be connected by all-JJTLs. However, the utilization of passive transmission lines (PTLs) looks more promising for long-range connections. It was shown that PTLs are suitable for high-density routing with low cross-talk allowing fast data transfer and precision timing design of the circuits \cite{PTL1ASC2020,PTL2ASC2020}. The cells can be even connected by PTLs only \cite{PTL3ASC2020}, though with a relatively large width of PTLs this option poses certain limits on the integration density.

While the considered bistable JJs can be readily modeled with simple two-JJ cells \cite{Suppl}, the full advantage from their utilization can be gathered with their fabrication as a solid heterostructure. Therefore, the development of corresponding fabrication methods suitable for large scale integrated circuits would be of significant benefit.

\section{Conclusion}

In summary, we considered different options for the implementation of the basic storing element of superconducting circuits. We argued that the use of inductors in conventional designs presents technological problems for circuit deep scaling. The utilization of bistable Josephson junction as a storing element looks promising in this context. In this case, magnetic flux is not used as the physical representation of information. This eliminates the requirements posed on inductors including the possibility to completely get rid of them. We presented the design methodology of the circuits without inductors based on bistable Josephson junctions. The methodology was used in the design of various basic cells like controlled oscillator analogous to conventional SFQ-to-DC converter, T flip-flop, D flip-flop, NDRO, logical NOT, and half adder. A more complex benchmark logic circuit of an 8-bit parallel adder was designed as well. Working margins of the all-JJ NDRO cell with small parasitic inductances at low-speed test exceed $\pm 20\%$ in simulations. The cell was designed using just a couple of values of the JJ damping coefficient, which is favorable for its implementation. The design of the half adder cell has shown that the total number of JJs in the all-JJ circuits can be less than in the standard design. We concluded that the proposed utilization of bistable Josephson junctions is promising in the aspects of the elimination of the need for quantizaing inductance, possible simplification of the all-JJ circuit schematics, and the reduction of total JJ count leading to space-efficiency. The search for technological routes for the bistable JJ fabrication and the development of information processing methods using circuits based on such junctions are urgent tasks in this area of research.

\begin{acknowledgments}
The authors are grateful to D. S. Holmes for fruitful discussions. The development of the all-JJ circuit design methodology was performed under the support of grant No. 20-12-00130 of the Russian Science Foundation. The review of the scalability of superconducting inductor was supported by the grant of the President of the Russian Federation MD-186.2020.8. Design of an 8-bit parallel adder was supported by RFBR grant No. 19-32-90208. V.I.R. acknowledges the Basis Foundation scholarship. N.V.K. and M.Yu.K. are grateful to the Interdisciplinary Scientific and Educational School of Moscow University ``Photonic and Quantum Technologies. Digital Medicine''.
\end{acknowledgments}

\bibliography{JJLM}

\end{document}



\title{Supplementary material to the paper ``Superconducting circuits without inductors based on bistable Josephson junctions''}

\author{I.~I.~Soloviev}
\email{isol@phys.msu.ru}
\affiliation{Lomonosov Moscow State
	University Skobeltsyn Institute of Nuclear Physics, 119991 Moscow,
	Russia} 
\affiliation{Dukhov All-Russia Research Institute of Automatics, Moscow 101000, Russia}
\affiliation{Physics Department of Moscow State University, 119991 Moscow, Russia}

\author{V.~I.~Ruzhickiy}
\affiliation{Lomonosov Moscow State
	University Skobeltsyn Institute of Nuclear Physics, 119991 Moscow,
	Russia}
\affiliation{Dukhov All-Russia Research Institute of Automatics, Moscow 101000, Russia}
\affiliation{Physics Department of Moscow State
	University, 119991 Moscow, Russia}

\author{S.~V.~Bakurskiy}
\affiliation{Lomonosov Moscow State
	University Skobeltsyn Institute of Nuclear Physics, 119991 Moscow,
	Russia} 
\affiliation{Dukhov All-Russia Research Institute of Automatics, Moscow 101000, Russia}
\affiliation{Moscow Institute of Physics and Technology, State
	University, 141700 Dolgoprudniy, Moscow region, Russia}

\author{N.~V.~Klenov}
\affiliation{Lomonosov Moscow State
	University Skobeltsyn Institute of Nuclear Physics, 119991 Moscow,
	Russia} 
\affiliation{Dukhov All-Russia Research Institute of Automatics, Moscow 101000, Russia}
\affiliation{Physics Department of Moscow State
	University, 119991 Moscow,
	Russia}

\author{M.~Yu.~Kupriyanov}
\affiliation{Lomonosov Moscow State
	University Skobeltsyn Institute of Nuclear Physics, 119991 Moscow,
	Russia} 

\author{A.~A.~Golubov}
\affiliation{Moscow Institute of Physics and Technology, State
	University, 141700 Dolgoprudniy, Moscow region, Russia}
\affiliation{Faculty of Science and Technology and MESA+ Institute of Nanotechnology, 7500 AE Enschede, The Netherlands}

\author{O.~V.~Skryabina}
\affiliation{Lomonosov Moscow State
	University Skobeltsyn Institute of Nuclear Physics, 119991 Moscow,
	Russia}
\affiliation{Moscow Institute of Physics and Technology, State
	University, 141700 Dolgoprudniy, Moscow region, Russia}

\author{V.~S.~Stolyarov}
\affiliation{Moscow Institute of Physics and Technology, State
	University, 141700 Dolgoprudniy, Moscow region, Russia}
\affiliation{Dukhov All-Russia Research Institute of Automatics, Moscow 101000, Russia}

\date{\today}

\maketitle

In this supplementary material, we investigate the Josephson Junction (JJ) Current-Phase Relation (CPR) conditions necessary to provide JJ bistability. Further, we consider cell models for the bistable $2\phi$- and $\phi$-JJ, and present details of our numerical calculations of circuits based on bistable junctions. 

\section{Josephson junction bistability condition}

The current-phase relation of a JJ possessing bistability has a second harmonic:
\begin{equation} \label{isCPR}
    i_s(\phi) = A\sin(\phi) + B\sin(2\phi),
\end{equation}
where the superconducting current, $i_s$, and the amplitudes, $A$, $B$, of the CPR components are real normalized to a reference critical current, $I_c$. While $A$ is assumed to be positive, $B$ may be positive or negative.
The corresponding Energy-Phase Relation (EPR) of the JJ is
\begin{equation}
    E(\phi) = A[1 - \cos(\phi)] + (B/2)[1 - \cos(2\phi)],
\end{equation}
where the energy, $E$, is normalized by the Josephson energy, $E_{J0} = I_c\Phi_0/2\pi$, $\Phi_0$ is the magnetic flux quantum.
The second derivative of this EPR is
\begin{equation}
    E'' = A\cos(\phi) + 2 B \cos(2\phi).
\end{equation}
The existence of an additional minimum on the EPR at $\phi = \pi$ corresponds to the positive sign of the second derivative, $E''(\pi) > 0$, which leads to the condition
\begin{equation}
    B > A/2.
\end{equation}
This condition can be fulfilled only for positive values of $B$. 

\begin{figure}[t]
	\includegraphics[width=1\columnwidth,keepaspectratio]{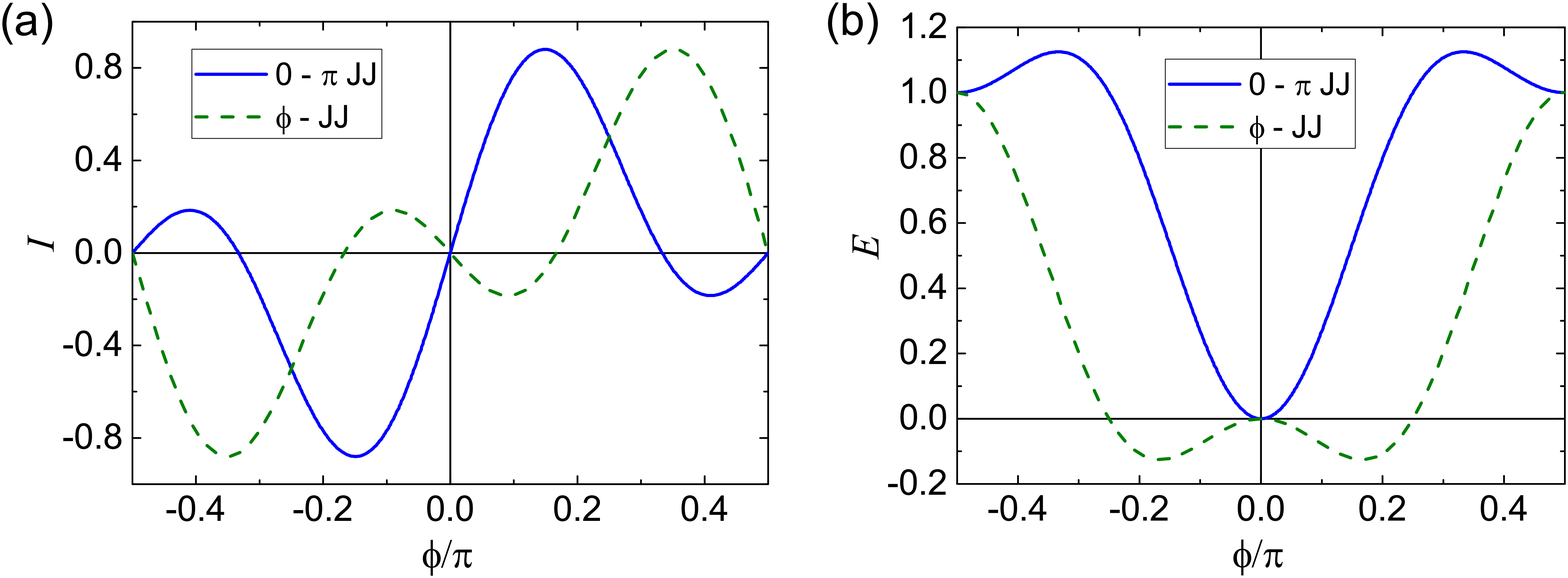}
	\caption{CPR (a) and EPR (b) of $0 - \pi$ (solid line) and $\phi$ (dashed line) Josephson junction. $A = |B| = 0.5$.
   } \label{SFig1}
\end{figure}

The decrease of $B$ down to the negative values finally gives rise to the origin of two minima on the EPR located symmetrically around zero. This corresponds to the negative sign of the second derivative of the EPR at zero, $E''(0) < 0$, that leads to
\begin{equation}
    B < - A/2.
\end{equation}
This condition can be fulfilled only for negative values of $B$. Therefore, the bistability of a JJ corresponds to the condition:
\begin{equation}
    |B| > A/2,
\end{equation}
where the sign of the second harmonic amplitude, $B$, determines the type of bistable JJ: $0 - \pi$ JJ \cite{0piJJ1,0piJJ2,0piAPhiJJ} for $B > 0$ and $\phi$-JJ \cite{PhiJJ1,PhiJJ2,PhiJJ3,PhiJJ4mem,PhiJJ5,0piAPhiJJ} for $B < 0$. In the particular case where $B > 0$ and $A = 0$, the JJ CPR has doubled periodicity so that such JJ is called $2\phi$-JJ \cite{T,RHM,PBRB,BSPR,Ryaz2phi}. The examples of CPR and EPR of $0 - \pi$ and $\phi$-JJ for $A = |B| = 0.5$ are shown in Fig.~\ref{SFig1}.

\section{Cell models of bistable Josephson junctions}

\begin{figure}[t]
	\includegraphics[width=0.3\columnwidth,keepaspectratio]{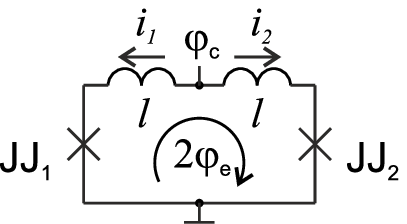}
	\caption{Schematic of a cell model for $\phi$- and $2\phi$-JJ.
   } \label{SFig2}
\end{figure}

In this section, we present cell models for $2\phi$-JJ and $\phi$-JJ. Both models are two-JJ cells with small but finite inductance, see Fig.~\ref{SFig2}. Here the currents $i_{1,2}$ are normalized on $I_c$, while the inductance is normalized by the Josephson inductance, $l = 2\pi I_c L/\Phi_0$. The CPR of the cell shown in Fig.~\ref{SFig2} is the dependence of the current sum, $i_+ = i_1 + i_2$, on the cell phase, $\varphi_c$.

The cell model of a $2\phi$-JJ is a circuit with two 0-JJs and a half flux quantum applied to the cell, $\varphi_e = \pi/2$, where $\varphi_e = \pi\Phi_e/\Phi_0$ is the external phase, $\Phi_e$ is the external magnetic flux. For each shoulder of this circuit, the cell phase, $\varphi_c$, equals to the sum of the phase drop on the JJ, $\phi_{1(2)}$, plus the phase drop on the inductance, $i_{1(2)}l = l\sin(\phi_{1(2)})$ (where we put the critical currents of the JJs equal to $I_c$) if $\varphi_e$ is zero. Under the assumption that the external flux evenly introduces the phase increase into the loop, the phase constraints are formalized by a system of equations,
\begin{equation} \label{sysphie}
    \phi_{1,2} + l\sin(\phi_{1,2}) = \varphi_c \pm \varphi_e.
\end{equation}
If the inductance is vanishing, $l = 0$, then $\phi_{1,2} = \varphi_c \pm \varphi_e$ so that at $\phi_e = \pi/2$ the cell has zero critical current. For $l \ll 1$, we assume that
\begin{equation} 
    \phi_{1,2} = \varphi_c \pm \varphi_e + \delta_{1,2}, \label{phi12}
\end{equation}
where $\delta_{1,2} \ll 1$. Substitution of (\ref{phi12}) into the system (\ref{sysphie}) leads to
\begin{equation}
    \delta_{1,2} \approx - \frac{l\sin(\varphi_c \pm \varphi_e)}{1 + l\cos(\varphi_c \pm \varphi_e)}. \label{delta}
\end{equation}
The current dependencies $i_{1,2}(\varphi_c)$ can be found with the substitution of (\ref{delta}) into (\ref{phi12}):
\begin{equation}
    i_{1,2} = \sin(\varphi_c \pm \varphi_e  + \delta_{1,2}) \approx \sin(\varphi_c \pm \varphi_e) - \frac{l}{2}\sin(2[\varphi_c \pm \varphi_e]). \label{i12}
\end{equation}
The total current through the cell versus the cell phase at $\phi_e = \pi/2$ is
\begin{equation}
    i_+ \approx l\sin(2\varphi_c), \label{CPRphie}
\end{equation}
correspondingly. The obtained CPR (\ref{CPRphie}) shows that the considered cell is the $2\phi$-JJ model with an effective critical current equal to $l$.

The $\phi$-JJ cell model is a circuit (Fig.~\ref{SFig2}) with 0- and $\pi$-JJ. The external phase is zero. The system of equations formalizing the phase constraints is now reads
\begin{equation} \label{sysphi}
    \phi_{1,2} \pm l\sin(\phi_{1,2}) = \varphi_c,
\end{equation}
where JJ$_1$ is 0-JJ and JJ$_2$ is $\pi$-JJ. The amendments to the cell phase in (\ref{phi12}) are 
\begin{equation}
    \delta_{1,2} \approx \mp \frac{l\sin(\varphi_c)}{1 \pm l\cos(\varphi_c)}. \label{deltai}
\end{equation}
The currents, $i_{1,2}$, are 
\begin{equation}
    i_{1,2} = \pm \sin(\varphi_c + \delta_{1,2}) \approx \pm \sin(\varphi_c) - \frac{l}{2}\sin(2\varphi_c), \label{i12}
\end{equation}
correspondingly. This results in the total current through the cell,
\begin{equation}
    i_+ \approx -l\sin(2\varphi_c). \label{CPRphi}
\end{equation}
The obtained expression (\ref{CPRphi}) corresponds to the $\phi$-JJ CPR with vanishing first harmonic. The cell effective critical current is equal to $l$ again.

Note that the inductors of the cell (Fig.~\ref{SFig2}) can be substituted for 0-JJs. In this case, e.g., the system (\ref{sysphi}) has the following form,
\begin{equation} \label{sysw0JJs}
    \phi_{1,2} + \phi_{l1,2} = \varphi_c,
\end{equation}
where $\phi_{l1,2}$ are the Josephson phases of the substituting 0-JJs. The currents flowing through the cell are
\begin{equation} \label{i0JJs}
    i_{1,2} = \pm \sin(\phi_{1,2}) = i_{cl}\sin(\phi_{l1,2}),
\end{equation}
where $i_{cl}$ is the critical current of the substituting 0-JJs. If $i_{cl} \gg 1$ then the equation (\ref{i0JJs}) leads to
\begin{equation} \label{phi0JJs}
    \phi_{l1,2} \approx \pm i_{cl}^{-1}\sin(\phi_{1,2}).
\end{equation}
The substitution of the expressions (\ref{phi0JJs}) into (\ref{sysw0JJs}) reproduces the system (\ref{sysphi}) but with $l$ substituted for $i_{cl}^{-1}$. Therefore, the effective critical current of the cell model here is $i_{cl}^{-1}$.

\section{Numerical modeling}

\subsection{Basic equations}

The considered cell models of bistable JJs are useful in the experimental implementation of the circuit prototypes since there is no proven technological method for the fabrication of such JJs with reproducible parameters nowadays. At the same time, the utilization of cell models in a simulation of complex circuits is painful because each JJ should be presented as a composite block of elements. Instead of this, we are modeling the circuits using equations including the second harmonic in the JJ CPR.

In the frame of the RCSJ model \cite{RSJC}, the total current through the JJ is the sum of three components: the superconducting one (\ref{isCPR}), the resistive, $i_n = (\Phi_0/2\pi)\dot{\phi}/I_cR$, and the capacitive, $i_{cap} = (\Phi_0/2\pi)\ddot{\phi}C/I_c$, where $R$ is the junction resistance in the normal state, and $C$ is the JJ capacitance. By introducing the characteristic frequency, $\omega_c = (2\pi/\Phi_0)I_cR$, and the plasma frequency, $\omega_p = \sqrt{(2\pi/\Phi_0)I_c/C}$, one can present the total current as follows:
\begin{equation} \label{JJcur}
i = A \sin\phi + B \sin 2\phi + \omega_c^{-1}\dot{\phi} + \omega_p^{-2}\ddot{\phi}.
\end{equation}
It is convenient to normalize the time in (\ref{JJcur}) to one of the frequencies, $\omega_c$, $\omega_p$. If it is normalized to the plasma frequency, equation (\ref{JJcur}) reads,
\begin{equation} \label{JJcurn}
i = A \sin\phi + B \sin 2\phi + \alpha\dot{\phi} + \ddot{\phi},
\end{equation}
where $\alpha = \omega_p/\omega_c$ is the JJ damping coefficient which is mainly defined by the JJ resistive shunt used to damp ``plasma oscillations'' occurring after the junction switching.

Complex digital circuits require perfect synchronization of their parts. This can be achieved with the fabrication of Josephson junctions having equal time characteristics. Standard superconducting technological process \cite{HighJc} is  characterized by a certain critical current density, $j_c$, and JJ specific capacitance, $c$, so that the junction critical current is $I_c = j_c a_{JJ}$ and the capacitance is $C = c a_{JJ}$, where $a_{JJ}$ is the junction area. Thus, the plasma frequency, $\omega_p = \sqrt{(2\pi/\Phi_0)j_c/c}$, is fixed for all JJs on a wafer. This allows us to apply the same normalization of time in the description of all JJs in the circuit. In this way, each JJ has only two basic parameters: the critical current, and the damping coefficient. 

Note that the extraction of parasitic inductance and capacitance from layout is very important for accurate calculations. If the bulk geometry of the junction produces parasitic capacitance, it is accounted for in calculation by the appropriate connection of the shunt capacitance in parallel to the JJ. This can be done by the introduction of a factor to the second derivative of phase in (\ref{JJcurn}).

\subsection{Calculation method}

Only Josephson junctions are present in our schematics. The circuit simulation is based on numerical calculations of a system of differential equations of the 2nd order of the type (\ref{JJcurn}), accordingly. This system is a consequence of the combination of the loop phase constraints, and Kirchoff’s current laws. At the first step, we convert the initial system to the system of equations of the 1st order: $y = \dot{\phi}$, $\dot{y} = \ddot{\phi}$. This doubles the number of equations. Then we use one-step explicit Runge-Kutta methods of the 4th and 5th order to find a solution at each time step. 

Our results can be reproduced in three ways: (i) by using the approach described above, (ii) by utilization of standard simulation tools with substitution of the bistable JJs by their cell models, or (iii) by the introduction of the second CPR harmonic into the open-code simulation software.

\subsection{Details of designs and simulations}

The proposed circuits are based on the basic block designed in accordance with the methodology presented in \cite{piJJDR}. 

The parameters of the JJs connected to the ground terminal in all-JJTLs (see Fig.~4(c), main text) are $A = 1, ~ \alpha = 1$, and the ones of interconnecting JJ are $A = 0.7, ~ \alpha = 1$. The bias current applied to all-JJTLs is $i_b = I_b / I_c = 0.75$.

The parameters of the Josephson junctions corresponding to different modes of operation of the basic block (see Fig.~4(d), main text) with and without controlled readout are summarized in Table~I. The parameters of the input valve junction, $J_v$, for the modes 1 - 4 considered in the main text (1 - terminator, 2 - transmission line, 3 - digital frequency divider, and 4 - oscillator) are shown in Fig.~5(a), main text. 

Note that while simple qualitative consideration supposes rectangular shapes for the areas of modes 1 and 2 in Fig.~5(a) (main text), their actual shapes correspond to the shunting of the current flowing through $J_m$ by $J_l$.

\begin{singlespace}
\begin{table}[t]
\centering
	\newcolumntype{Y}{>{\hsize=0.8\hsize\linewidth=\hsize\centering\arraybackslash}X}
	\setlength{\extrarowheight}{3pt}
	\begin{tabularx}{\textwidth}{>{\hsize=3.4\hsize\linewidth=\hsize}X *{12}Y}
		\multicolumn{13}{l}{TABLE I. Parameters of JJs for different modes of the basic block operation.}\\
		\hline \hline ~ & $A_{in}$ & $\alpha_{in}$ & $B_v$ & $\alpha_v$ &
		$B_m$ & $\alpha_m$ & $A_l$ & $\alpha_l$ & $A_{out}$ & $\alpha_{out}$
		& $A_r$
		& $\alpha_r$ \\
		\hline
		Modes 1 - 4 of the basic block & 1 & 1 &  \multicolumn{2}{c}{Fig.~5(a)} & 1 & 2 & 0.5 & 1 & 1 & 1 & - & - \\ [1ex]
		\hline
		Controlled readout & 1 & 5 & 0.1 & 5 & 0.5 & 1 & 0.5 & 1/3.5$^{a}$ & 2 & 2 & 1$\backslash$1.5$^{b}$ & 1 \\
		\hline \hline
		\multicolumn{13}{l}{$^{a}$ NDRO / DRO.}\\
		\multicolumn{13}{l}{$^{b}$ Without $\backslash$ with output inversion.}
	\end{tabularx}
\end{table}
\end{singlespace}

In mode 4, the increase in $J_v$ potential energy in the total potential energy of the basic block (proportional to $B_v$) results in high deviation of Josephson phases of 0-JJs from zero (at $J_m$ phase about $\pi + 2\pi n$). Their potential energies become higher, accordingly. Starting from a certain threshold (which is seen as a vertical boundary between the modes 3 and 4 in Fig.~5(a), main text) the output junction $J_{out}$ is near the top of its potential energy barrier. Leakage of the interface all-JJTL bias current into $J_{out}$ provides its periodic switching. 

For the implementation of the NDRO cell (Fig.~6(a), main text) we take parameters of interconnecting junctions, $J_v$, $J_l$, inherited from the mode 3 of the basic block operation. The bias current of $J_{out}$ is $i_b = 1.5$. Switching to DRO mode requires an increase in the $J_l$ damping, as is explained in the main text. Note that the damping coefficient of the input junction $J_{in}$ is also increased. This ensures that the read $2\pi$ phase change wave is terminated by switching $J_v$ after switching $J_m$.

The read data inversion can be performed by a simple swap of $J_r$, $J_{out}$ connection order, see Fig.~6(f), main text. In this case, $J_m$ and $J_{out}$ are connected to the ground through $J_r$, so that the stationary distribution of the currents and dynamics of the Josephson junctions are changed. The critical current of $J_r$ is increased here to preserve the operation of the JJs, accordingly. Another difference in the operation is that $J_r$ Josephson phase increase by $2\pi$ leads to the generation of the phase change wave propagating toward the output all-JJTL and the cell input. The last wave is terminated by switching of $J_v$ leaving $J_{in}$ phase unchanged.

The parameters of the half-adder (HA) JJs (see Fig.~8(a), main text) are presented in Table~II. The bias currents, $i_{bc}$, $i_{bm}$, $i_{bs}$, are 0.75, 0.6, and 1.3, correspondingly.

\begin{singlespace}
\begin{table}[t]
\centering
	\newcolumntype{Y}{>{\hsize=0.8\hsize\linewidth=5\hsize\centering\arraybackslash}X}
	\setlength{\extrarowheight}{3pt}
	\begin{tabularx}{\textwidth}{>{\hsize=3\hsize\linewidth=\hsize}Y *{9}Y}
		\multicolumn{9}{l}{TABLE II. Parameters of half-adder Josephson junctions.}\\
		\hline \hline ~ & $J_{c(a,b)}$ & $J_c$ & $J_{lc}$ & $J_{v(a,b)}$ &
		$J_m$
		& $J_{ls}$ & $J_s$ & $J_{rs}$ \\
		\hline
		$A(B^{\text{~a}})$ & 0.1 & 1 & 0.4 & 0.1 & 1 & 0.1 & 2 & 0.9 \\
		$\alpha$ & 1 & 1 & 1 & 3 & 1 & 4& 3 & 3 \\
		\hline \hline \multicolumn{9}{l}{\footnotesize$^\text{a}$ $B$ for $J_{v(a,b)}$ and $J_m$.}\\
	\end{tabularx}
\end{table}
\end{singlespace}

In the proposed HA cell, the confluence buffer (CB) connecting data branches to $J_m$ is assembled from sequential pairs of $2\phi$-JJs: $J_{v(a)}$, $J_m$ and $J_{v(b)}$, $J_m$. The incoming $2\pi$ phase changes are transformed into $\pi$ phase drops on each pair of these junctions. If only one phase change wave comes, e.g., from line $a$, the phases of the interconnecting junction, $J_{v(a)}$, and the main junction, $J_m$, become increased by $\pi$. At the same time, the phase of the interconnecting junction in the neighboring data branch, $J_{v(b)}$, decreases by $\pi$. If the second phase change wave comes (from line $b$), the previously gained $\pi$ phase increase/decrease on $J_{v(a)}$/$J_{v(b)}$ becomes canceled, and the $J_m$ phase is increased up to $2\pi$. The same effect takes place if the phase change waves initially come from both data branches simultaneously.

Just a single $2\phi$-JJ ($J_m$) is sufficient to provide data storage. This is in contrast to RSFQ HA where insensitivity to the data delay comes at the cost of doubling the SFQ storage circuit \cite{RSFQHA}. Thus, the number of JJs in the presented HA schematic is even smaller than in the RSFQ counterpart despite the substitution of inductors for JJs.

\bibliography{JJLM}